\documentclass{article}

\PassOptionsToPackage{numbers}{natbib}

\usepackage[preprint]{neurips_2026}
\usepackage{xcolor}
\usepackage{colortbl}
\usepackage{algorithm}
\usepackage{algorithmic}
\usepackage{amssymb}
\usepackage{enumitem}
\usepackage{tcolorbox} 
\tcbuselibrary{listings,breakable,skins}
\usepackage{enumitem}
 
\usepackage{listings}
\usepackage{multirow}
\usepackage{booktabs}
\usepackage{wrapfig}

\usepackage[normalem]{ulem}
\useunder{\uline}{\ul}{}
\usepackage{subcaption}
\usepackage{caption}
\usepackage{bm}
\usepackage[british, american]{babel}

\definecolor{softred}{RGB}{200,120,120}
\definecolor{softgreen}{RGB}{120,170,120}

\usepackage{tikz}
\usetikzlibrary{shapes.geometric}

\definecolor{mygreen}{RGB}{0,150,0}
\definecolor{myred}{RGB}{200,0,0}

\definecolor{citecolor}{HTML}{0071BC}
\definecolor{linkcolor}{HTML}{ED1C24}

\usepackage[pagebackref=true,breaklinks=true,colorlinks,bookmarks=false,citecolor=citecolor,linkcolor=linkcolor,urlcolor=gray]{hyperref}

\usepackage[utf8]{inputenc}
\usepackage[T1]{fontenc}
\usepackage{url}
\usepackage{graphicx}
\usepackage{booktabs}
\usepackage{amsfonts}
\usepackage{nicefrac}
\usepackage{microtype}
\usepackage{amsmath}
\usepackage[capitalize]{cleveref}
\crefname{section}{Sec.}{Secs.}
\Crefname{section}{Sec.}{Secs.}
\crefname{subsection}{Sec.}{Secs.}
\Crefname{subsection}{Sec.}{Secs.}
\crefname{subsubsection}{Sec.}{Secs.}
\Crefname{subsubsection}{Sec.}{Secs.}
\crefname{figure}{Fig.}{Figs.}
\Crefname{figure}{Fig.}{Figs.}
\crefname{table}{Tab.}{Tabs.}
\Crefname{table}{Tab.}{Tabs.}
\crefname{equation}{Eq.}{Eqs.}
\Crefname{equation}{Eq.}{Eqs.}
\crefname{algorithm}{Alg.}{Algs.}
\Crefname{algorithm}{Alg.}{Algs.}
\usepackage{booktabs}

\usepackage{pifont}
\definecolor{ygreen}{RGB}{0,140,0}
\definecolor{yred}{RGB}{200,30,30}

\definecolor{linkpink}{RGB}{210, 80, 140}

\title{GeoVolDiff: Taming 3D Geological Volumes with Latent Diffusion}

\author{
\begin{tabular}{c}
\textbf{Qi Pang \quad Hongling Chen\thanks{Corresponding author} \quad Jinghuai Gao} \\[0.3em]
\normalfont Xi'an Jiaotong University \\
\texttt{pangjiutian@gmail.com, chenhongling@xjtu.edu.cn, jhgao@xjtu.edu.cn} \\[0.3em]
{\hypersetup{urlcolor=linkpink}
Code: \url{https://github.com/PangJiutian/geovoldiff}}
\end{tabular}
}

\begin{document}

\maketitle

\begin{abstract}
\vspace{-.5em}
Deep learning has become a prevailing paradigm across a wide range of geophysical applications. Yet most existing studies concentrate on methodological refinements—novel network architectures, physics-informed constraints, or task-specific loss functions—while paying comparatively little attention to a more fundamental challenge of any data-driven approach: the availability and representativeness of high-quality training data. This limitation is especially pronounced in geophysics. Unlike computer vision, which benefits from large-scale, well-curated benchmarks such as ImageNet, comparably abundant and reliably labelled geophysical data are prohibitively expensive to acquire and, in most field settings, lack accessible ground-truth supervision.
To alleviate this data deficiency, we propose \textbf{GeoVolDiff}, a generative framework for three-dimensional geological volumes. It comprises three coupled stages: (i) constructing a foundational training corpus through physics-based forward simulation; (ii) training a Latent Diffusion Model (LDM) to capture the statistical distribution of 3D geological structures; and (iii) synthesizing diverse, structurally plausible volumes at scale for downstream geophysical tasks. We examine the utility of the synthesized data on a representative downstream task, seismic impedance inversion. Without incorporating any additional physical or geological prior, inversion networks pre-trained exclusively on synthesized data attain competitive performance on both synthetic and field datasets, indicating that data synthesised by the generative model can serve as an effective surrogate for costly field-acquired labels.

\end{abstract}

\section{Introduction}
Over the past decade, neural networks have substantially challenged
classical model-driven pipelines across a broad range of geophysical
tasks, including seismic inversion~\citep{wang2020well,wu2021deep,
chen2025unsupervised,pang2025iterative}, noise
attenuation~\citep{meng2022seismic, 11007646, meng2025posterior}, geobody segmentation~\citep{gao2021channelseg3d, 11175213}, structural interpretation~\citep{wu2019faultseg3d, gao2021fault}, full-waveform inversion~\citep{zhang2021deep, zhang2020adjoint, zhang2022regularized}, and seismic super-resolution~\citep{yang2025seismic, liu2023improving, li2026high}. The underlying principle
is consistent: a network with millions of tunable parameters selects
a solution from an enormous model space, and the training
data is what constrains that search—telling the network which answers
are geologically and physically plausible. The quality
and volume of the training data is therefore not merely an implementation detail but
a fundamental determinant of what any data-driven model can learn.
Yet most methodological effort has gone into architectural
innovation—physics-informed loss terms, attention mechanisms, and
self-supervised objectives—while the question of \textit{where
sufficient, representative training data come from} has received
comparatively little attention.

In geophysics, this question is especially difficult to answer. Ground
truth is largely inaccessible: borehole logs provide reliable labels, but they are sparse, one-dimensional
samples—a handful of vertical profiles scattered across an
otherwise unobserved three-dimensional volume. Field seismic
acquisition is expensive and logistically demanding, and no
full-resolution model of the actual subsurface exists, because the
earth cannot be excavated for
verification. The network is therefore left to generalise across
most of the volume with no direct supervision.
This data scarcity is the central problem we address.

A natural first response is to generate training data synthetically
through physics-based forward modelling~\citep{merrifield2022synthetic, wu2020building}—constructing structural
geological models and computing their seismic responses from
prescribed subsurface parameters. This approach is physically
consistent and provides reliable ground-truth labels, making it
a sound foundation for a training corpus. Its critical limitation,
however, lies in scalability and structural diversity: conventional
geomodelling demands extensive domain expertise and complex parameter
tuning and is highly labour-intensive, making it impractical to
produce the breadth of structural variety that downstream deep
learning models require. Even automated pipelines such as that
of~\citep{wu2020building} still rely on hand-designed rules and
careful parameter tuning. What is needed is a mechanism that
takes this modest, physically grounded seed corpus and automatically
expands it into a large, structurally diverse dataset—and this is
precisely where generative modelling enters.

Recent advances in computer vision provide compelling evidence that such distribution-learning and data-generation capabilities are achievable. Latent Diffusion Models (LDMs)~\citep{rombach2022high}, which learn to reverse a gradual noising process in a compressed latent space, capture the statistical structure of natural images remarkably well and generate large numbers of diverse, high-fidelity samples not seen during training. Extensions such as text-to-image diffusion~\citep{zhang2023adding,mou2024t2i,li2023gligen} further enable controllable generation through textual and structural conditioning, and video diffusion models~\citep{blattmann2023stable,wan2025wan,blattmann2023align,gao2025seedance} show that diffusion-based frameworks extend to high-dimensional structured data while preserving long-range spatial and temporal coherence. Crucially, these models do not merely memorise; they
generalise the underlying data distribution. This
motivates a direct analogy: \textit{if a diffusion model can
learn the distribution of natural images and generate
photorealistic scenes that have never existed, can it learn
the statistical distribution of geology and generate
realistic, never-before-seen 3D subsurface volumes?}

Building upon our preliminary study on unconditional diffusion-based geological volume synthesis~\citep{pang2026scaling}, we develop \textbf{GeoVolDiff},
a three-stage generative pipeline for 3D geological volume
synthesis. A physics-based forward simulator first builds
a compact but physically grounded seed corpus of labelled 3D
volumes; an LDM then learns their structural distribution;
and large-scale synthesis from the trained model supplies
the training pool for downstream tasks. We validate this
synthesis-driven approach on seismic impedance inversion,
and find that networks trained exclusively on GeoVolDiff-generated
data—without any externally imposed geological prior or
low-frequency model—achieve competitive accuracy on held-out
synthetic benchmarks and transfer to real field datasets.
These results position synthesis-driven data generation as
a practical, scalable strategy for alleviating the
labelling bottleneck in 3D geophysical deep learning.

\section{Methods}
\label{sec:method}

\begin{figure}[t]
    \centering
    \includegraphics[width=1\linewidth]{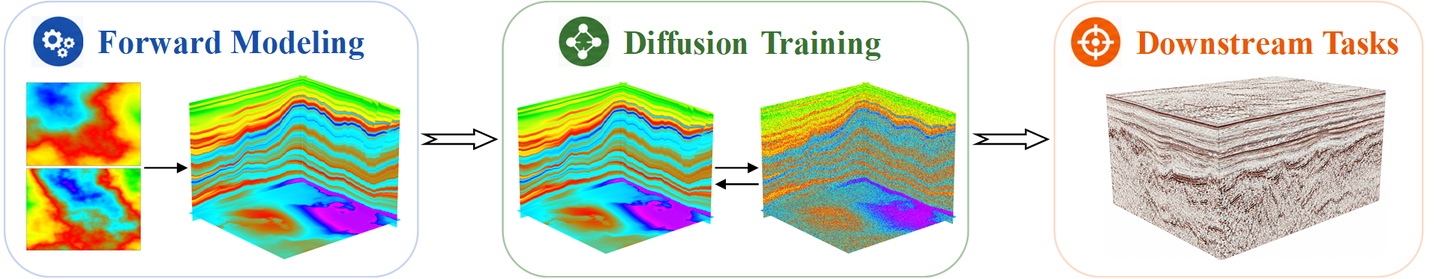}
    \caption{\textbf{Overview of the GeoVolDiff framework.} Three sequential stages: forward simulation of 3D geological volumes with paired condition labels, training of a 3D latent diffusion model, and large-scale data synthesis for downstream geophysical tasks.
    }
    \label{fig:pipeline}
    \vspace{-1em}
\end{figure}

We present \textbf{GeoVolDiff}, our framework
for generating three-dimensional geological volumes via latent diffusion
modelling. As illustrated in Fig.~\ref{fig:pipeline}, it
consists of three stages: (1) constructing a high-quality training
dataset through physics-based forward simulation; (2) training a
Latent Diffusion Model to learn the prior distribution of geological
volumes; and (3) using the trained model to synthesize diverse
data for pre-training downstream geophysical tasks.

\subsection{3D Forward Simulation Framework}
\begin{figure}[t]
    \centering
    \includegraphics[width=0.90\linewidth]{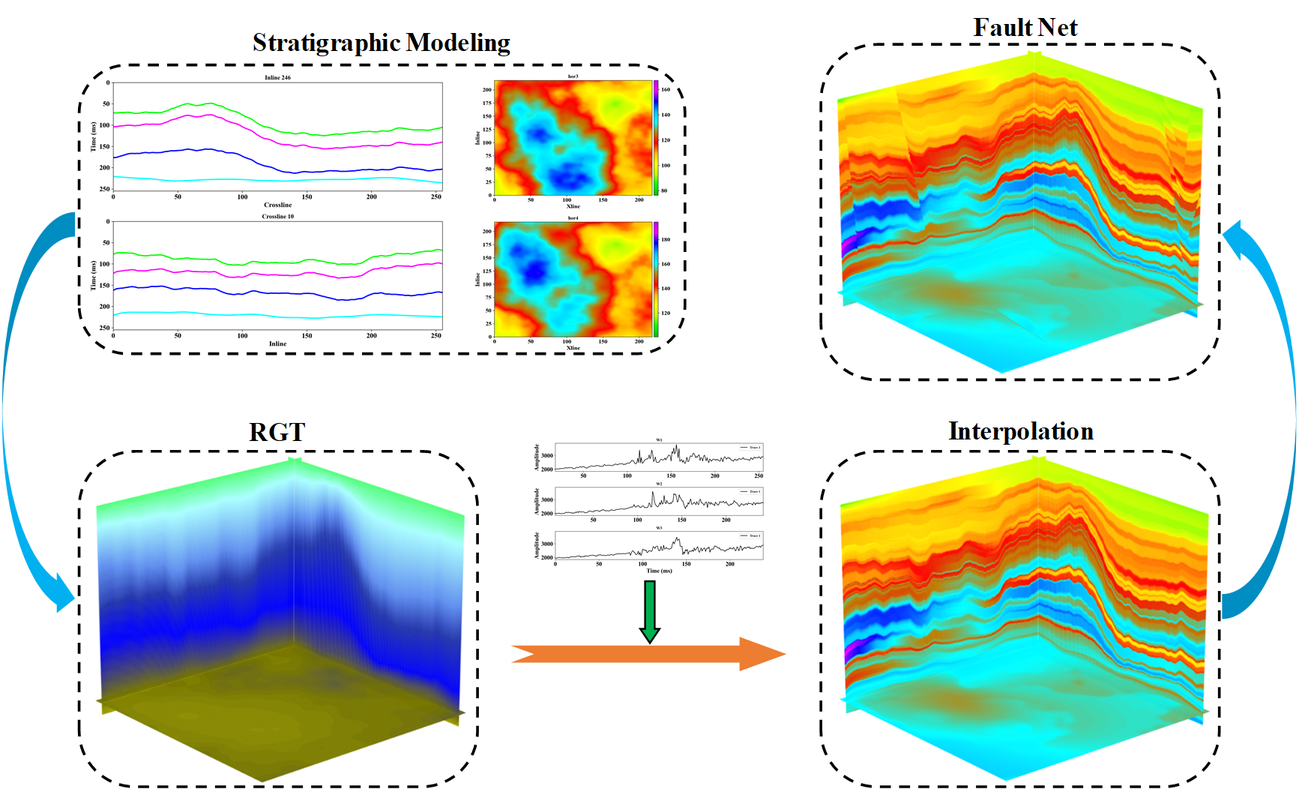}
    \caption{\textbf{Parameterised forward-simulation workflow.} Stratigraphic modelling, RGT volume construction, attribute interpolation along the RGT scaffold, and fault-network embedding, yielding the geological volume $\mathbf{M}$ together with paired labels.
    }
    \label{fig:forward_modeling}
    \vspace{-1em}
\end{figure}

An ideal training corpus for 3D geological volume generation should satisfy three requirements simultaneously: (i) \textit{geophysical plausibility}—the samples obey first-order stratigraphic deposition rules and structural-deformation principles; (ii) \textit{structural diversity}—a broad range of stratigraphic architectures, fault geometries, and depositional configurations; and (iii) \textit{data sufficiency}—a sample count sufficient for training contemporary deep generative models. Conventional geological modelling pipelines depend heavily on manual interpretation, which is labour-intensive and difficult to scale to the dataset sizes relevant for generative learning. We therefore design a fully parameterised, automated simulation workflow that produces 3D geological attribute models, as illustrated in Fig.~\ref{fig:forward_modeling}.

\vspace{-0.5em}
\paragraph{Stratigraphic modeling.}
The geometry of the initial stratigraphic sequence is generated with fractal Perlin noise~\citep{perlin1985image}, which yields spatially continuous, smoothly varying surfaces that emulate the gentle undulations and lateral facies transitions typical of sedimentary strata. Each horizon $z_i(x,y)$ is parameterised as a scaled noise field added to a reference depth, and stacking the horizons yields the full stratigraphic framework $\mathcal{H}$. By controlling the noise frequency and amplitude independently per layer, the workflow systematically samples geometric variability across realisations.

\vspace{-1em}
\paragraph{Relative geological time modeling.}
Given the stratigraphic framework $\mathcal{H}$, each horizon is assigned a monotonically increasing relative geologic time (RGT) value $t_i$ that encodes the depositional order from oldest (bottom) to youngest (top). Linear interpolation between successive horizons then produces a continuous 3D RGT volume, yielding a layer-conformable coordinate frame $\mathcal{S}$ that enforces stratigraphic consistency in the downstream attribute-interpolation step.

\vspace{-1em}
\paragraph{Attribute interpolation.}
With the stratigraphic scaffold $\mathcal{S}$ available, sparse well-log traces are introduced as soft attribute anchors and propagated into the full 3D domain by depth-slice-wise spatial interpolation guided by $\mathcal{S}$, producing an initial attribute model $\mathbf{M}_0 \in \mathbb{R}^{N_x \times N_y \times N_z}$. The well logs here serve purely as carriers of petrophysical values rather than as strict geological constraints; they may come from field measurements, synthetic logs, or manually specified profiles, which affords considerable flexibility in data sourcing. To enrich structural and numerical diversity across realisations, controlled stochastic perturbations are applied to both well locations and attribute amplitudes, and a blocking operation converts continuous logs into piecewise-constant profiles that more faithfully reflect the layered heterogeneity of subsurface formations.

\vspace{-1em}
\paragraph{Fault network embedding.}
Fault structures are embedded into $\mathbf{M}_0$ following the structural-modelling formulation of Wu et al.~\citep{wu2020building}, producing the final attribute model $\mathbf{M} \in \mathbb{R}^{N_x \times N_y \times N_z}$. The fault module supports parameterised generation of flower-structure fault systems—both positive and negative variants—as well as multi-fault configurations with prescribed distributions of strike, dip, throw, and lateral extent. The number of faults per realisation is drawn uniformly from a predefined range, ensuring systematic variability in structural complexity. Importantly, the workflow produces, as a natural by-product, paired condition labels—3D fault masks—that are directly reused as conditioning signals for conditional latent diffusion modelling in the next stage.

\subsection{3D Latent Diffusion Model}

Generative modelling of 3D geological volumes poses two fundamental challenges. \textit{First}, geological volumes exhibit strong structural coupling across all three spatial directions—depth, inline, and crossline—so the model must represent spatial dependencies jointly along the three axes rather than independently. \textit{Second}, their high dimensionality leads to substantial GPU-memory consumption, making diffusion modelling in the native voxel space computationally inefficient and memory-prohibitive in practice. To address both issues, we adopt a two-stage 3D Latent Diffusion Model: a 3D Variational Autoencoder (3D-VAE) is first trained to compress the volumes into a low-dimensional latent space, and a conditional diffusion process is then learned entirely within that space.

\vspace{-0.5em}
\paragraph{3D Variational Autoencoder}

\begin{figure}[h]
    \centering
    \includegraphics[width=1\linewidth]{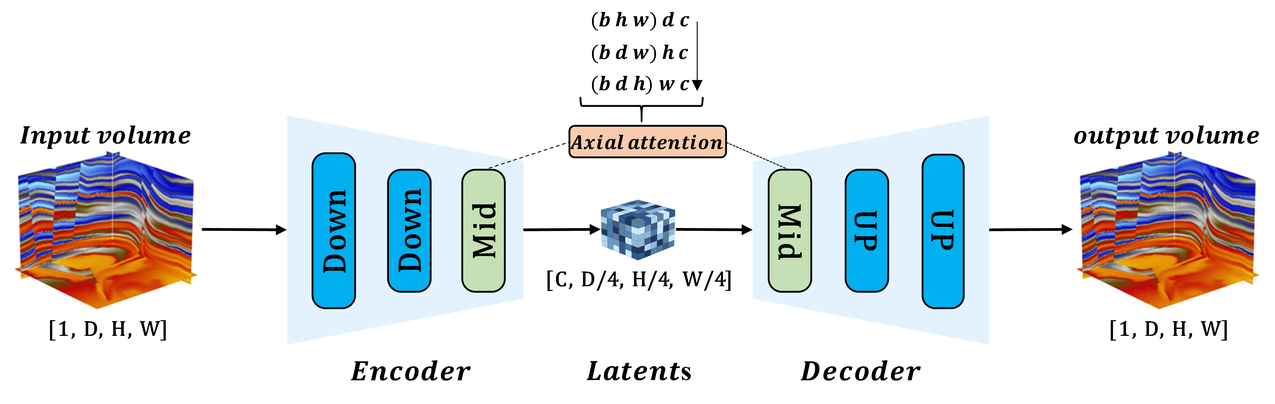}
    \caption{\textbf{Architecture of the 3D-VAE.} 3D-convolutional encoder--decoder with axial-attention modules at the bottleneck.
    }
    \label{fig:vae}
    \vspace{-0.5em}
\end{figure}

A high-fidelity VAE is essential within the LDM framework: it compresses high-resolution volumetric data into a low-dimensional latent representation while regularising the latent distribution toward an isotropic Gaussian prior, thereby supporting stable training of the downstream diffusion model. To accommodate the strong inter-axis structural coupling of geological data, we use 3D convolutions as the basic building block of both the encoder and the decoder, as illustrated in Fig.~\ref{fig:vae}.

To capture long-range spatial dependencies at the VAE bottleneck while remaining tractable for large seismic volumes, we adopt a memory-efficient variant of axial attention. Given a feature tensor $\mathbf{F} \in \mathbb{R}^{B \times C \times D \times H \times W}$ (batch, channels, depth, height, width), attention is computed in parallel along each spatial axis and the outputs are averaged:
\begin{equation}
\mathbf{F}^{\prime} = \frac{1}{3}\left(
    \mathcal{A}_D\!\left(\mathbf{F}\right) +
    \mathcal{A}_H\!\left(\mathbf{F}\right) +
    \mathcal{A}_W\!\left(\mathbf{F}\right)
\right)
\label{eq:axial_attn_vae}
\end{equation}
where $\mathcal{A}_D(\cdot)$, $\mathcal{A}_H(\cdot)$, and
$\mathcal{A}_W(\cdot)$ denote axial self-attention
along the depth, inline, and crossline dimensions,
respectively. This parallel formulation reduces memory overhead relative to full 3D self-attention, and independent per-axis attention suffices to capture the salient spatial structure needed for reconstruction.

The VAE is trained with a composite objective combining a voxel-wise MSE term, a KL-divergence regularisation term $D_{\mathrm{KL}}$, and a perceptual similarity term $\mathcal{L}_{\mathrm{LPIPS}}$, following 
the formulation of~\citep{rombach2022high}:
\begin{equation}
    \mathcal{L}_{\mathrm{VAE}} = 
    \lambda \left\| \mathbf{x} - \hat{\mathbf{x}} \right\|_2^2
    + \beta\, D_{\mathrm{KL}}\!\left(q_\phi(\mathbf{z}\mid\mathbf{x}) 
    \,\|\, p(\mathbf{z})\right)
    + \mathcal{L}_{\mathrm{LPIPS}}\!\left(\mathbf{x},\, \hat{\mathbf{x}}\right)
    \label{eq:vae_loss}
\end{equation}
where $\mathbf{x}\in\mathbb{R}^{H\times W\times D\times C}$ denotes the input three-dimensional geological volume, $\hat{\mathbf{x}}$ its reconstruction, and $\mathbf{z}$ the corresponding latent variable encoded by the VAE. The perceptual term $\mathcal{L}_{\mathrm{LPIPS}}$ enforces structural fidelity beyond pure voxel-wise reconstruction, encouraging the network to preserve the spatial patterns and geological textures of subsurface formations. Because the underlying VGG-based feature extractor~\citep{zhang2018unreasonable} accepts only 2D inputs, we uniformly sample cross-sections along each of the three principal axes, compute $\mathcal{L}_{\mathrm{LPIPS}}$ independently on the three orthogonal slice families, and average over the three orientations. This multi-orientation slicing promotes isotropic perceptual fidelity and partially mitigates the inherently 2D nature of LPIPS in a 3D setting.

\vspace{-0.5em}
\paragraph{3D Conditional Latent Diffusion Model}
\begin{figure}[h]
    \centering
    \includegraphics[width=1\linewidth]{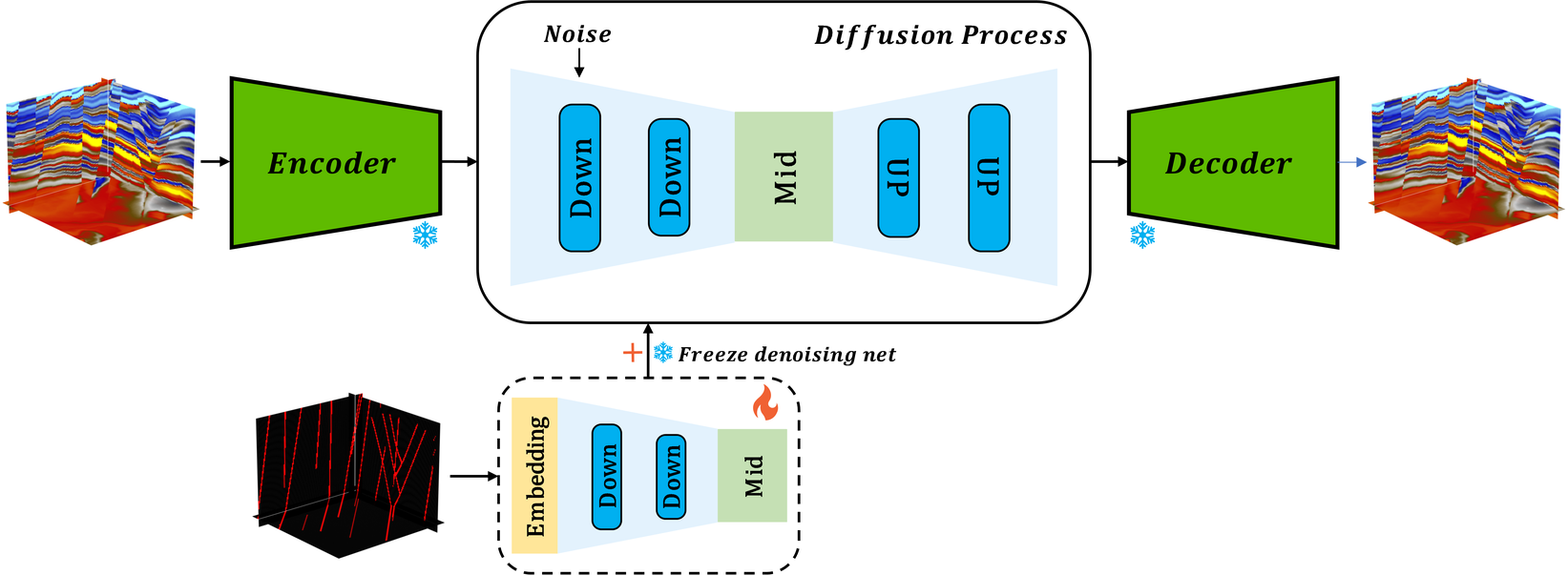}
    \caption{\textbf{3D conditional latent diffusion model.} Denoising network operating in the VAE latent space, with ControlNet branch injecting fault-mask conditioning through trainable residual connections.
    }
    \label{fig:ldm}
    \vspace{-0.5em}
\end{figure}

With the VAE parameters frozen, the diffusion model operates entirely in the latent space $\mathbf{z} = \mathcal{E}(\mathbf{x})$ produced by the encoder $\mathcal{E}(\cdot)$~\citep{rombach2022high}. The denoising network is a UNet built primarily from 3D convolutional blocks, which enables multi-scale feature representation while preserving volumetric spatial correlations throughout the reverse diffusion process. To further model long-range spatial dependencies, 
sequential axial attention is inserted at every encoder stage, decoder 
stage, and bottleneck layer, operating along the depth, inline, and 
crossline axes:
\begin{equation}
    \mathbf{F}^{\prime}
    =
    \mathcal{A}_W\!\left(
    \mathcal{A}_H\!\left(
    \mathcal{A}_D(\mathbf{F})
    \right)\right)
\end{equation}
Unlike the parallel formulation used in the VAE, the sequential design lets each axis attend over context already refined by the preceding axis, which is critical for maintaining coherent geological structure during iterative denoising. Following the standard formulation, the network is trained to reverse a $T$-step Markovian forward noising process by predicting the noise injected at each timestep~\citep{ho2020denoising}, with the objective:
\begin{equation}
    \mathcal{L}_{\mathrm{LDM}} = 
    \mathbb{E}_{\mathbf{z}, \boldsymbol{\epsilon} \sim \mathcal{N}(\mathbf{0}, \mathbf{I}), t}
    \left[ \left\| \boldsymbol{\epsilon} - 
    \boldsymbol{\epsilon}_\theta\!\left(\mathbf{z}_t, t, \mathbf{c}\right) 
    \right\|_2^2 \right]
    \label{eq:ldm_loss}
\end{equation}
where $\mathbf{z}_t$ denotes the latent state at diffusion timestep $t$ and $\mathbf{c}$ is the conditioning signal. This latent-space formulation substantially reduces the per-step computational cost relative to native-space 3D diffusion, while inheriting the compactness of the VAE latent representation.

To enable geometrically controllable generation, the fault labels produced as by-products of the forward-simulation pipeline are introduced as structural conditioning signals through a ControlNet branch~\citep{zhang2023adding}. The branch takes a binary 3D fault mask $\mathcal{F} \in \{0,1\}^{N_x \times N_y \times N_z}$ as input and injects fault geometry into the denoising network through trainable residual connections, while the backbone diffusion model is kept frozen. As illustrated in Fig.~\ref{fig:ldm}, this design provides explicit spatial control over fault location and geometry at generation time without perturbing the geological distribution captured by the backbone.

\subsection{Downstream Task Validation}

\begin{figure}[h]
    \centering
    \includegraphics[width=1\linewidth]{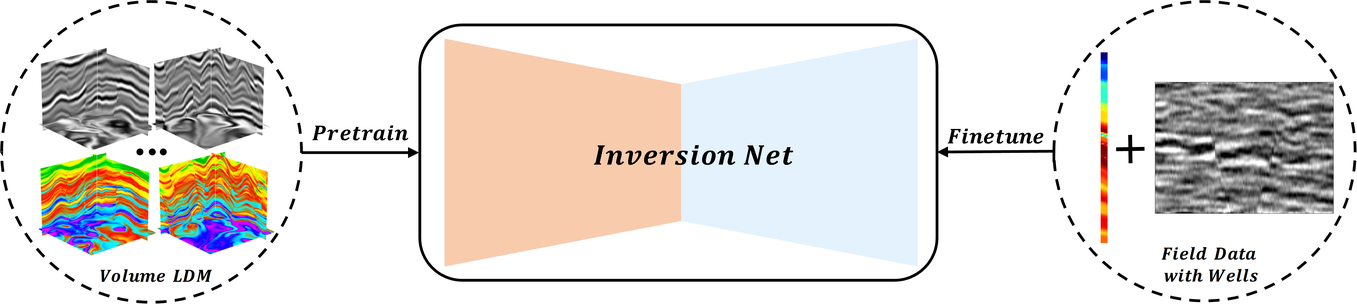}
    \caption{\textbf{Pretrain--finetune pipeline for downstream seismic impedance inversion.} Impedance synthesis with GeoVolDiff, paired-data construction via 1D convolutional forward modelling, pre-training on synthetic data, and fine-tuning with field well logs.
    }
    \label{fig:dm_inv}
    \vspace{-1em}
\end{figure}

To examine the practical utility of the data synthesized by GeoVolDiff, we use a task-driven validation protocol in which the generated volumes serve as pre-training data for a downstream geophysical task. We adopt seismic acoustic-impedance inversion as a representative case, structured as a three-stage pipeline (Fig.~\ref{fig:dm_inv}).

In Stage 1, acoustic-impedance volumes are synthesized by the trained GeoVolDiff model. In Stage 2, the corresponding synthetic seismic data are produced by 1D convolutional forward modelling along the time axis, giving paired \textit{(seismic, impedance)} samples; a simple UNet-based inversion network is then pre-trained end-to-end on this fully synthetic corpus to learn the seismic-to-impedance mapping without any real field data. In Stage 3, the pre-trained network is fine-tuned using a small number of real well-log labels together with the corresponding field seismic traces, adapting the pre-learned representations to the target field domain.
\section{Experiments}
\label{sec:exp}

\subsection{Forward Simulation Results}

\begin{figure}[h]
    \centering
    \includegraphics[width=1\linewidth]{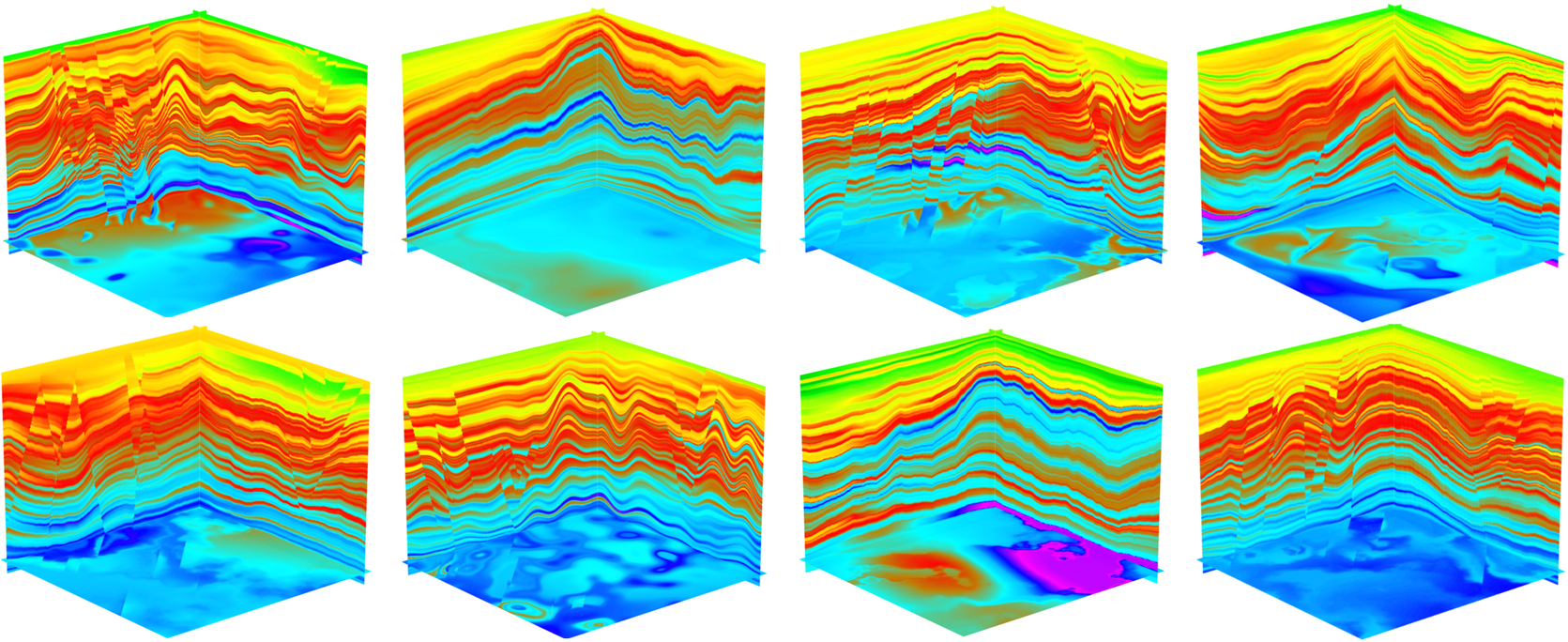}
    \caption{\textbf{Representative 3D geological volumes produced by the forward-simulation workflow.}
    }
    \label{fig:forward_results}
    \vspace{-1em}
\end{figure}

To provide a training corpus for the diffusion model, the forward-simulation workflow generates a structurally diverse dataset of 3D geological volumes. The dataset comprises 40 volumes at a native resolution of $256^3$ voxels; representative samples are shown in Fig.~\ref{fig:forward_results}.

We deliberately maintain a resolution gap between simulation and training: forward simulation is performed at $256^3$ voxels, whereas the diffusion model is trained at $128^3$ voxels owing to GPU-memory constraints. The aim is to suppress a well-known sampling-induced aliasing artifact: simulating directly at $128^3$ on a regular grid produces jagged, discontinuous boundaries along geological interfaces—most notably steeply dipping strata and fault planes—because the spatial sampling rate is too low to resolve the underlying geometry. Simulating at $256^3$ instead renders these interfaces with smooth, geometrically faithful boundaries; $128^3$ sub-volumes are then randomly cropped from the high-resolution volumes during training, which preserves structural fidelity while acting as a natural data-augmentation mechanism.

\subsection{3D Latent Diffusion Model}

\paragraph{3D VAE Reconstruction.}
\begin{figure}[h]
    \centering
    \includegraphics[width=1\linewidth]{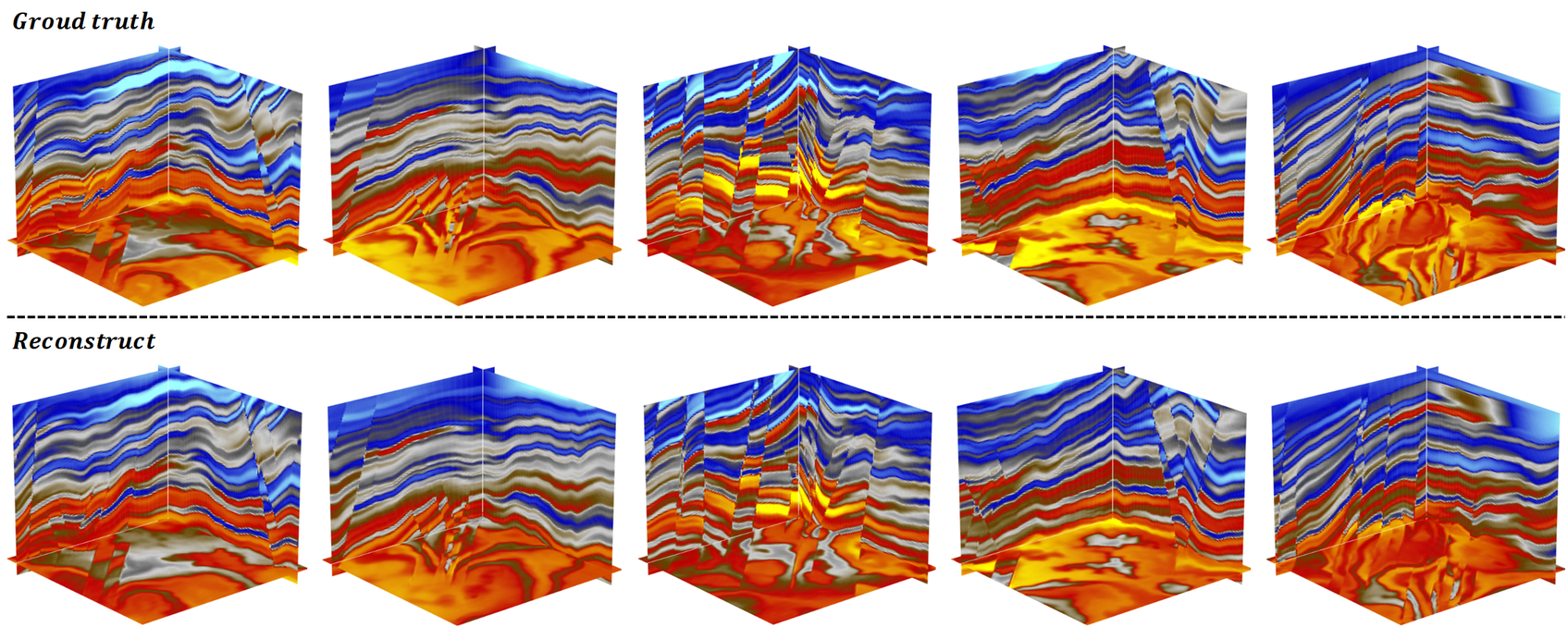}
    \caption{\textbf{3D-VAE reconstruction on out-of-training-set volumes.} Each pair shows the ground-truth volume and its encode--decode reconstruction.
    }
    \label{fig:vae_res}
    \vspace{-1em}
\end{figure}
The forward-simulated volumes are first randomly cropped into $128^3$ sub-volumes and then augmented by depth-axis flipping, in-plane rotation within the inline–crossline plane, and amplitude scaling and shifting, yielding a final VAE training set of 5{,}000 sub-volumes at $128^3$ resolution.

Reconstruction fidelity is evaluated on volumes generated independently by forward simulation and then passed through the trained encoder–decoder pipeline (Fig.~\ref{fig:vae_res}). Overall, the reconstructions preserve the global geometry and visual appearance of the ground-truth volumes, with high structural consistency at the macroscopic scale.

Mild degradation is nonetheless observed at high-frequency structural features—most notably along fault boundaries and thin layers—appearing as slight blurring and amplitude deviation. We attribute this to two factors. \textit{First}, model capacity is limited by GPU memory, which constrains the network's ability to represent fine-scale structural detail. \textit{Second}, the perceptual loss $\mathcal{L}_{\mathrm{LPIPS}}$ is anisotropically effective across orientations: reconstructions on inline and crossline sections are generally coherent and boundary-smooth, whereas time slices show noise-like, boundary-discontinuous patterns that lie outside the training distribution of the VGG feature extractor, degrading reconstruction accuracy along the time (depth) axis.
Despite these residual artefacts, the reconstruction quality is sufficient to provide a well-structured latent space for downstream diffusion-model training, as the generation results in the next section confirm.

\vspace{-0.1em}
\paragraph{Conditional Diffusion Model Generation.}

Whereas the VAE is optimised for per-volume reconstruction fidelity, the diffusion model must represent a broader diversity of structural configurations across multiple spatial scales. Accordingly, at every training step the cropping window is sampled from $\{128, 192, 256\}$ voxels, and the cropped sub-volume—together with its paired condition labels—is resampled to the uniform working resolution of $128^3$. Combined with on-the-fly augmentation (rather than the offline augmentation used at the VAE stage), this multi-scale random-cropping strategy ensures that the network sees a distinct data realisation at every iteration, maximising effective training diversity. The model is trained with batch size $2$ for $5{,}000$ epochs ($100{,}000$ optimisation steps in total) using a cosine noise schedule.

\begin{figure}[h]
    \centering
    \includegraphics[width=1\linewidth]{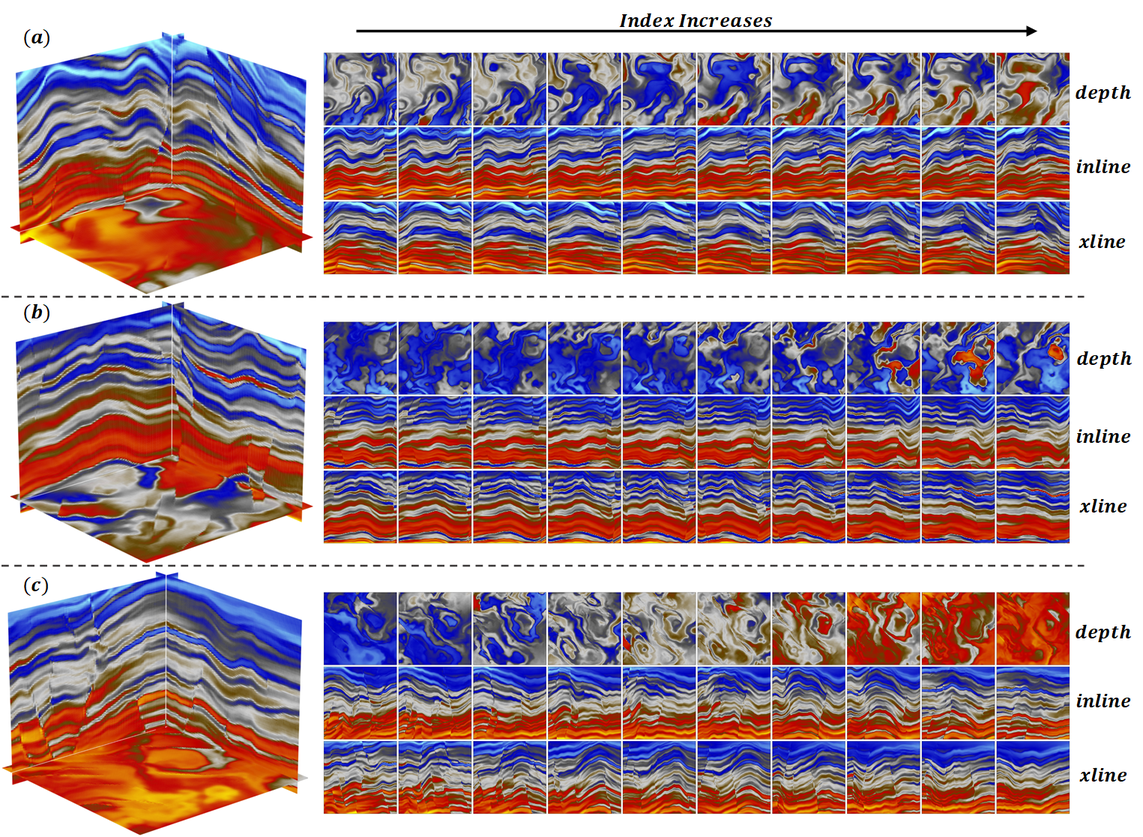}
    \caption{\textbf{Unconditional generation results.} Synthesized 3D volumes with inline, crossline, and time-slice cross-sections.
    }
    \label{fig:uncond_res}
    \vspace{-1em}
\end{figure}

\begin{figure}[h]
    \centering
    \includegraphics[width=1\linewidth]{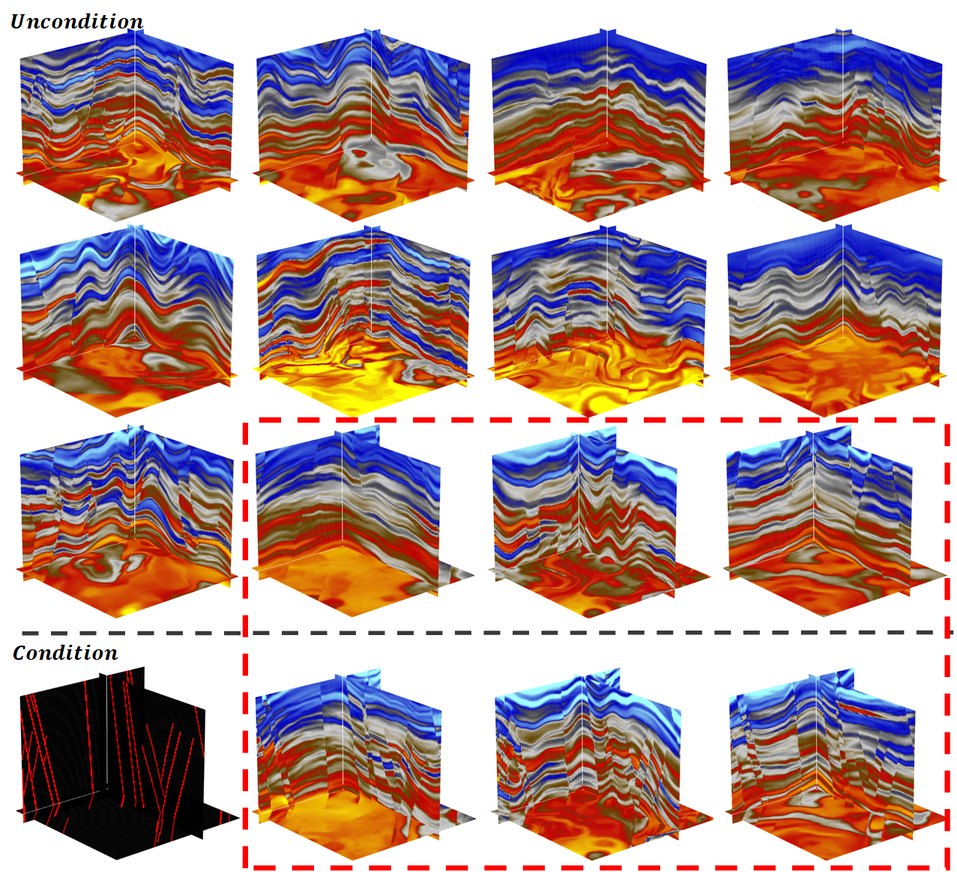}
    \caption{\textbf{Unconditional and fault-conditioned generation 
    results of GeoVolDiff.}
    The first three rows show unconditional samples, exhibiting diverse stratigraphic configurations and strong lateral continuity. The bottom row presents fault-conditioned generation: the leftmost volume displays the input fault mask, and the remaining three volumes show the corresponding generated 
    results. The red dashed box highlights volumes generated under fault conditioning, demonstrating precise geometric control over fault location and spatial extent.
    }
    \label{fig:cond_res}
    \vspace{-1em}
\end{figure}

We first examine the unconditional generation results in Fig.~\ref{fig:uncond_res}, which are consistent with those reported in our preliminary work~\citep{pang2026scaling}. The generated volumes exhibit strong lateral continuity and substantial structural diversity. The cross-sectional profiles evolve consistently along the three principal orientations, indicating that the diffusion model has captured the joint three-dimensional dependencies inherent in volumetric geological data rather than degenerating into axis-decoupled 2D generation. Additional unconditional samples, together with conditional-generation results, are provided in Fig.~\ref{fig:cond_res}. When 3D fault masks are introduced as structural conditioning signals, the generated fault geometry closely follows the constraints encoded in the input, showing that the ControlNet branch achieves effective geometric control without disrupting the geological priors captured by the backbone.

\subsection{Downstream Validation: Seismic Impedance Inversion}
To assess whether the synthesized data are both practically effective and geologically plausible, the GeoVolDiff-generated volumes are used directly as pre-training data for a seismic impedance inversion network, followed by lightweight fine-tuning on field seismic traces and well-log labels. To increase training diversity, 100 acoustic-impedance volumes of size $128^3$ are synthesized by unconditional sampling. The corresponding synthetic seismic data are obtained by trace-wise convolution with a 25\,Hz Ricker wavelet, with coherent noise added across a range of signal-to-noise ratios to approximate realistic acquisition conditions. To reduce computational cost, we validate in 2D: 32 inline slices are randomly extracted from each of the 100 volumes, yielding a pre-training set of $3{,}200$ 2D profiles at $128 \times 128$ resolution. The inversion network is pre-trained for $10{,}000$ iterations at an initial learning rate of $1 \times 10^{-3}$, with MSE as the sole training objective.

\vspace{-0.5em}
\paragraph{Synthetic data.}

\begin{figure}[h]
    \centering
    
    \begin{subfigure}[b]{0.32\linewidth}
        \centering
        \includegraphics[width=\linewidth]{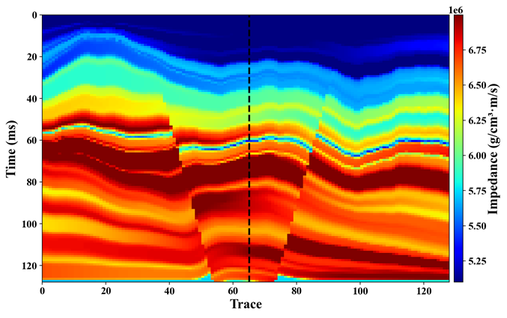}
        \caption{}
        \label{fig:a}
    \end{subfigure}
    \hfill
    \begin{subfigure}[b]{0.32\linewidth}
        \centering
        \includegraphics[width=\linewidth]{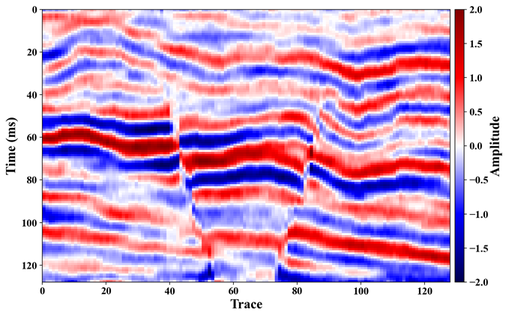}
        \caption{}
        \label{fig:b}
    \end{subfigure}
    \hfill
    \begin{subfigure}[b]{0.32\linewidth}
        \centering
        \includegraphics[width=\linewidth]{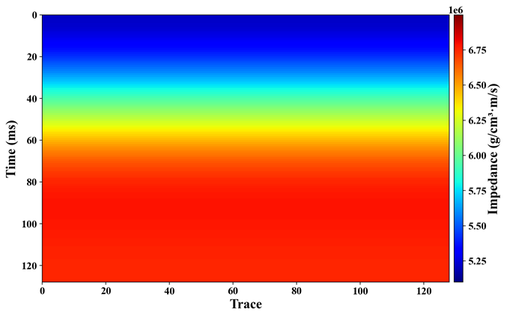}
        \caption{}
        \label{fig:c}
    \end{subfigure}

    \caption{
        \textbf{Synthetic test case.} \textbf{(a)} Ground-truth impedance model with well locations (dashed). \textbf{(b)} Synthetic seismic data at 10\,dB SNR. \textbf{(c)} Low-frequency background impedance used as initial model for USTNet.
    }
    
    \label{fig:dm_real_data}
    
    \vspace{-1em}
\end{figure}

Validation is first conducted on a synthetic case study, whose ground-truth impedance and corresponding seismic data are shown in Fig.~\ref{fig:dm_real_data}\subref{fig:a} and \subref{fig:b}, respectively, with dashed lines indicating well locations. The baseline is USTNet~\citep{pang2025iterative}, whose initial low-frequency impedance model is shown in Fig.~\ref{fig:dm_real_data}\subref{fig:c}. To probe the representations acquired during pre-training, the pre-trained network is first applied directly to the observed seismic data without fine-tuning. As is common in conventional learning-based inversion, USTNet relies on (i) a physics-consistent forward-loss term and (ii) a low-frequency background impedance model as the initial iterate, both needed to compensate for the scarcity of well-log supervision. The fine-tuning stage of the proposed framework, by contrast, requires neither an explicit forward loss nor a low-frequency initial model: through large-scale pre-training on synthesized data, the network has already implicitly encoded an approximate seismic-to-impedance mapping, making explicit physical constraints unnecessary during fine-tuning. This is particularly advantageous given that forward-loss terms depend on the accuracy of the assumed wavelet, an additional source of uncertainty that the proposed pre-training strategy effectively circumvents.

\begin{figure}[h]
    \centering

    \begin{subfigure}[b]{0.32\linewidth}
        \centering
        \includegraphics[width=\linewidth]{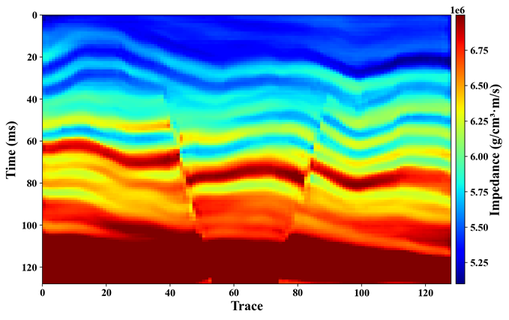}
        \caption{}
        \label{fig:a}
    \end{subfigure}
    \hfill
    \begin{subfigure}[b]{0.32\linewidth}
        \centering
        \includegraphics[width=\linewidth]{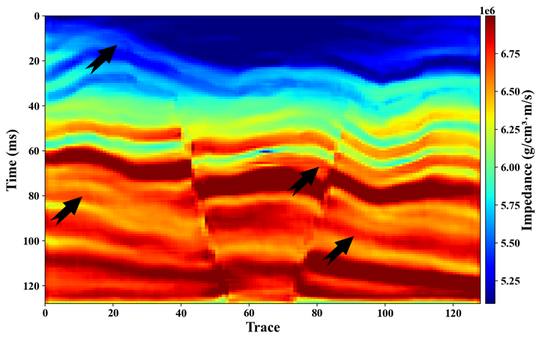}
        \caption{}
        \label{fig:b}
    \end{subfigure}
    \hfill
    \begin{subfigure}[b]{0.32\linewidth}
        \centering
        \includegraphics[width=\linewidth]{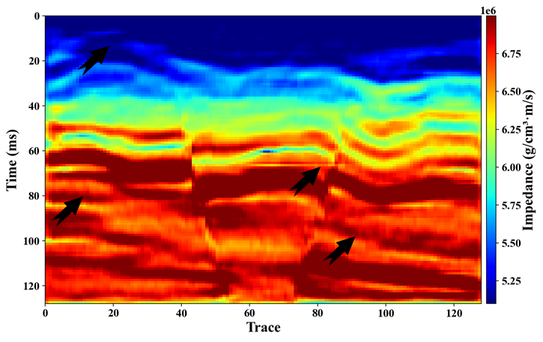}
        \caption{}
        \label{fig:c}
    \end{subfigure}
    \caption{
        \textbf{Inversion results on the synthetic case at 10\,dB SNR.} \textbf{(a)} Pre-trained network applied directly without fine-tuning. \textbf{(b)} Proposed pretrain--finetune framework. \textbf{(c)} USTNet baseline. Arrows mark the far-well region.
    }
    \label{fig:res_1}
    \vspace{-1em}
\end{figure}

\begin{figure}[h]
    \centering

    \begin{subfigure}[b]{0.48\linewidth}
        \centering
        \includegraphics[width=\linewidth]{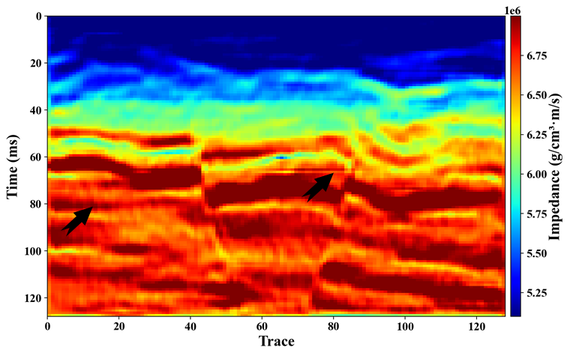}
        \caption{}
        \label{fig:a}
    \end{subfigure}
    \hfill
    \begin{subfigure}[b]{0.48\linewidth}
        \centering
        \includegraphics[width=\linewidth]{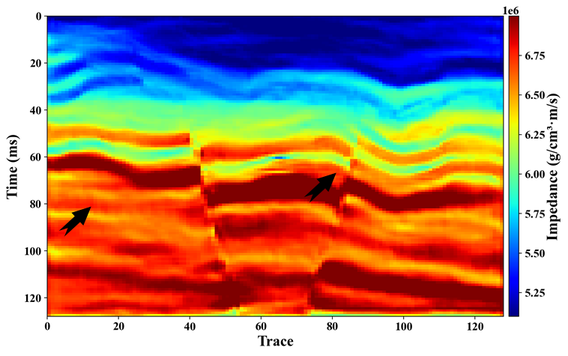}
        \caption{}
        \label{fig:b}
    \end{subfigure}

    \begin{subfigure}[b]{0.48\linewidth}
        \centering
        \includegraphics[width=\linewidth]{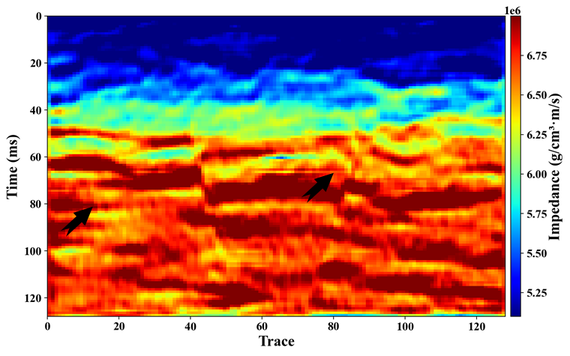}
        \caption{}
        \label{fig:c}
    \end{subfigure}
    \hfill
    \begin{subfigure}[b]{0.48\linewidth}
        \centering
        \includegraphics[width=\linewidth]{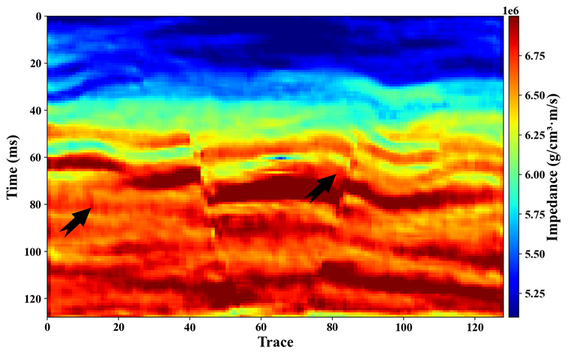}
        \caption{}
        \label{fig:d}
    \end{subfigure}

    \caption{\textbf{Inversion results under stronger noise.} \textbf{(a)} USTNet at 5\,dB. \textbf{(b)} Proposed framework at 5\,dB. \textbf{(c)} USTNet at 0\,dB. \textbf{(d)} Proposed framework at 0\,dB.}
    \label{fig:res_2}

    \vspace{-1em}
\end{figure}

As shown in Fig.~\ref{fig:res_1}\subref{fig:a}, even without additional prior constraints, the network recovers the overall structural configuration and principal spatial distribution of impedance from the seismic observations alone. This indicates that large-scale pre-training on GeoVolDiff-generated data equips the network with a meaningful prior over the macro-scale impedance distribution of the subsurface. Figs.~\ref{fig:res_1}\subref{fig:b} and \subref{fig:c} compare the proposed pretrain–finetune framework with USTNet. Both methods achieve comparable resolution near the wells. In the far-well region (black arrows), however, USTNet recovers only the overall impedance trend through its Transformer-based long-range modelling and forward-loss constraint, with insufficient lateral continuity, loss of fine structural detail, and visible inversion artefacts—signatures of limited extrapolation under sparse-label conditions. By contrast, the pre-trained inversion network preserves strong lateral continuity and well-defined stratigraphic boundaries across the entire profile, surpassing USTNet in both overall accuracy and fine-grained structural detail, while using neither a Transformer architecture nor a forward-loss term. 

The improvement in far-well extrapolation can be attributed to the 
transferable seismic-to-impedance prior acquired during pre-training. 
Nevertheless, the inevitable distribution mismatch between synthetic 
and field data precludes direct deployment of the pre-trained model. Fine-tuning with a small number of well logs aligns the learned representation with the field-data distribution, enabling the network to extend the acquired mapping beyond the conditioning wells and yield physically consistent impedance estimates across the entire seismic section. These results demonstrate that GeoVolDiff-generated volumes provide an effective pre-training corpus, enabling data-driven inversion networks to achieve robust generalisation under sparse well-log supervision. Additional comparisons at varying signal-to-noise ratios are provided in Fig.~\ref{fig:res_2}.

\vspace{-0.5em}
\paragraph{Field data.}
To further assess the geological plausibility and practical transferability of GeoVolDiff-generated data, we validate on two field datasets. \textit{Field dataset 1} exhibits gentle structural dip with laterally continuous reflectors, whereas \textit{Field dataset 2} is dominated by complex structures and thin-layer sequences. USTNet serves as the comparison baseline on Field dataset 1, and the inversion result provided by the dataset originator is taken as the reference on Field dataset 2.

\begin{figure}[h]
    \centering

    \begin{subfigure}[b]{0.48\linewidth}
        \centering
        \includegraphics[width=\linewidth]{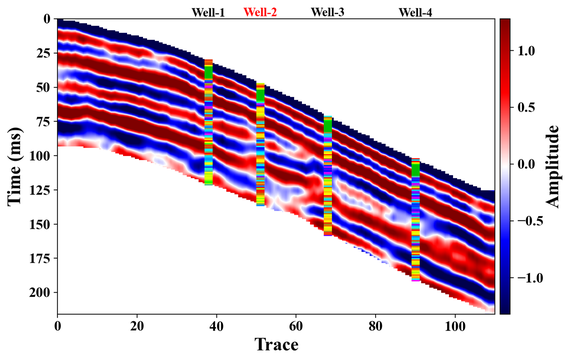}
        \caption{}
        \label{fig:a}
    \end{subfigure}
    \hfill
    \begin{subfigure}[b]{0.48\linewidth}
        \centering
        \includegraphics[width=\linewidth]{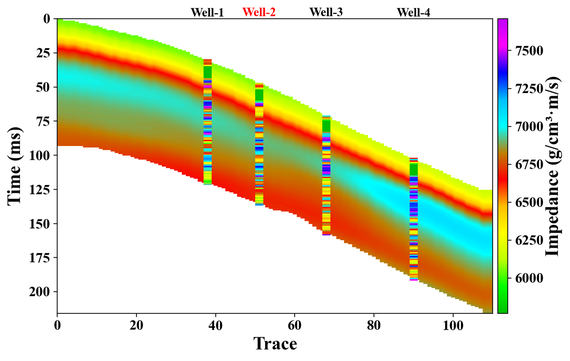}
        \caption{}
        \label{fig:b}
    \end{subfigure}
    \caption{\textbf{Field dataset 1.} \textbf{(a)} Observed seismic profile with well locations; Well-2 (red) is the validation well, Well-1/3/4 (black) are used for fine-tuning. \textbf{(b)} Low-frequency background impedance from well-log interpolation (used by USTNet only).}
    \label{fig:field_1}

    \vspace{-1em}
\end{figure}

\textit{Field dataset 1.}
Well-2 is held out as a blind well (red in Fig.~\ref{fig:field_1}\subref{fig:a}) for an unbiased quantitative assessment, while Well-1, Well-3, and Well-4 (black) provide well-log labels for fine-tuning and for the supervised constraints of USTNet. For USTNet, a low-frequency background impedance model is constructed by interpolating the available well logs and applying a low-pass filter to the result (Fig.~\ref{fig:field_1}\subref{fig:b}); the forward-loss term is parameterised with a wavelet estimated from the same well logs. For the proposed pre-trained network, no field-estimated wavelet is used during fine-tuning—a deliberate wavelet-mismatch scenario designed to probe the robustness of the pre-trained representations under wavelet-induced distribution shift.

\begin{figure}[h]
    \centering

    \begin{subfigure}[b]{0.48\linewidth}
        \centering
        \includegraphics[width=\linewidth]{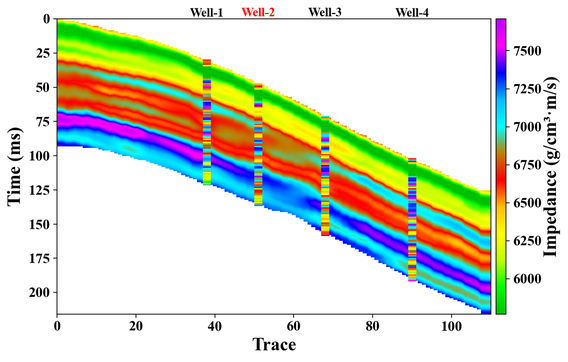}
        \caption{}
        \label{fig:a}
    \end{subfigure}
    \hfill
    \begin{subfigure}[b]{0.48\linewidth}
        \centering
        \includegraphics[width=\linewidth]{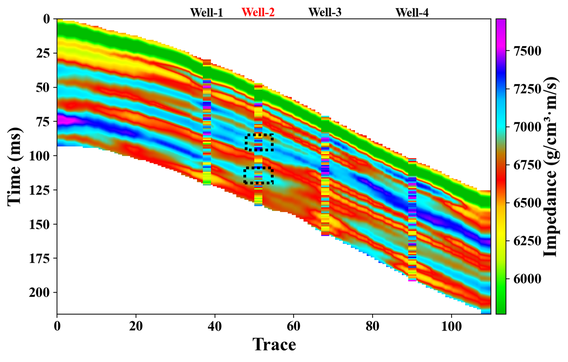}
        \caption{}
        \label{fig:b}
    \end{subfigure}

    \begin{subfigure}[b]{0.48\linewidth}
        \centering
        \includegraphics[width=\linewidth]{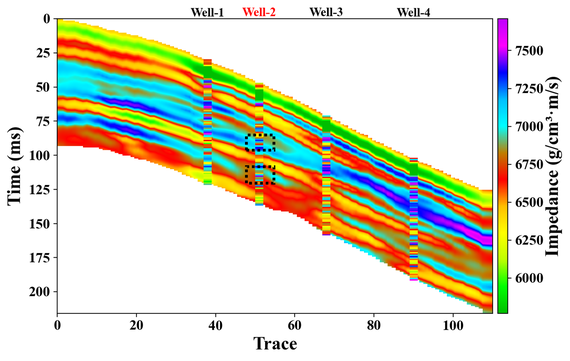}
        \caption{}
        \label{fig:c}
    \end{subfigure}

    \caption{\textbf{Inversion results on Field dataset 1.} \textbf{(a)} Pre-trained network without fine-tuning. \textbf{(b)} Proposed pretrain--finetune framework after well-log fine-tuning. \textbf{(c)} USTNet baseline with low-frequency background and field-estimated wavelet. Dashed box marks the vicinity of the blind well Well-2.}
    \label{fig:res_3}

    \vspace{-1em}
\end{figure}

Even before fine-tuning, the pre-trained network produces an impedance profile with reasonable lateral continuity, and the predicted values follow a monotonic shallow-to-deep increasing trend, consistent with the statistics of the pre-training dataset (Fig.~\ref{fig:res_3}\subref{fig:a}). This indicates that large-scale synthetic pre-training enables the network to capture a generalized seismic-to-impedance mapping. A pronounced bias nonetheless remains between the predicted impedance and the actual well-log values, suggesting that the amplitude-distribution mismatch between synthetic and real subsurface impedance is the primary factor limiting inversion accuracy. To bridge this gap, the pre-trained model is then fine-tuned with a small number of well logs. The fine-tuned results are compared with the baseline in Fig.~\ref{fig:res_3}\subref{fig:b} and \subref{fig:c}. Both methods produce laterally continuous profiles with well-resolved structural detail in the well-constrained regions, but the two pipelines draw on distinctly different priors: USTNet (Fig.~\ref{fig:res_3}\subref{fig:c}) is supervised by well-log labels and further augmented with a low-frequency background model constructed from well-log interpolation and a field wavelet used in the forward-loss term, both encoding strong macro-scale information about the impedance trend of the survey area. By contrast, the pretrain–finetune framework (Fig.~\ref{fig:res_3}\subref{fig:b}) relies solely on well-log labels during fine-tuning, without any supplementary low-frequency or wavelet-derived prior.

Near the blind well Well-2 (dashed box), USTNet (PCC: 0.8081) shows limited thin-layer resolution and fails to recover fine impedance variations within thin stratigraphic units. The pretrain–finetune framework (PCC: 0.8493), by contrast, achieves higher thin-layer discriminability in the same region and more faithfully delineates the lateral variation of thin-layer structures. We attribute this improvement to the lateral-continuity and thin-layer priors implicitly acquired during large-scale pre-training on GeoVolDiff-generated data, which give the network stronger lateral extrapolation and fine-detail reconstruction beyond the well-constrained region.

\begin{figure}[h]
    \centering

    \begin{subfigure}[b]{1\linewidth}
        \centering
        \includegraphics[width=\linewidth]{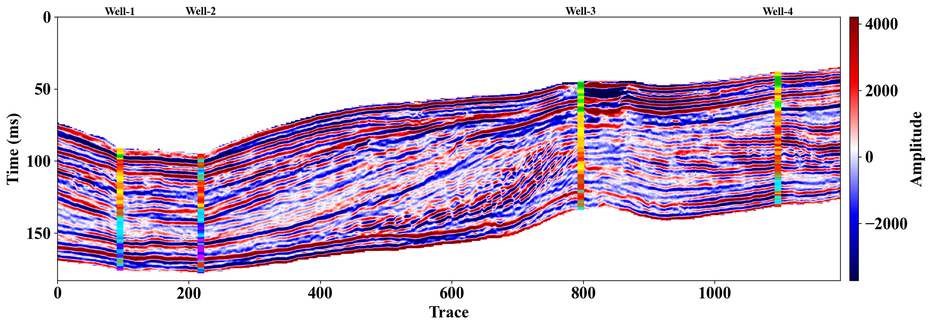}
        \caption{}
        \label{fig:a}
    \end{subfigure}

    \begin{subfigure}[b]{1\linewidth}
        \centering
        \includegraphics[width=\linewidth]{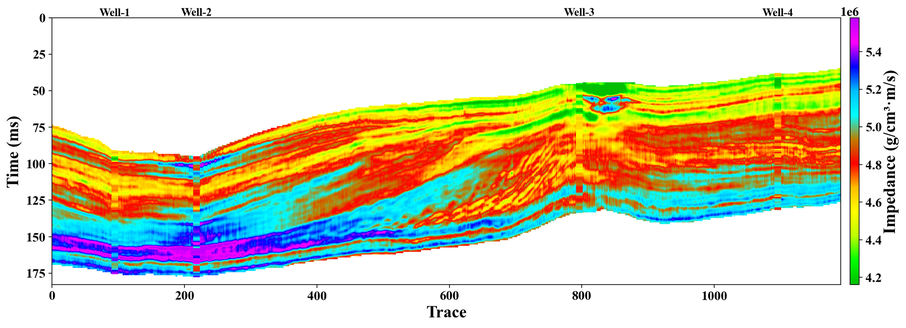}
        \caption{}
        \label{fig:b}
    \end{subfigure}

    \begin{subfigure}[b]{1\linewidth}
        \centering
        \includegraphics[width=\linewidth]{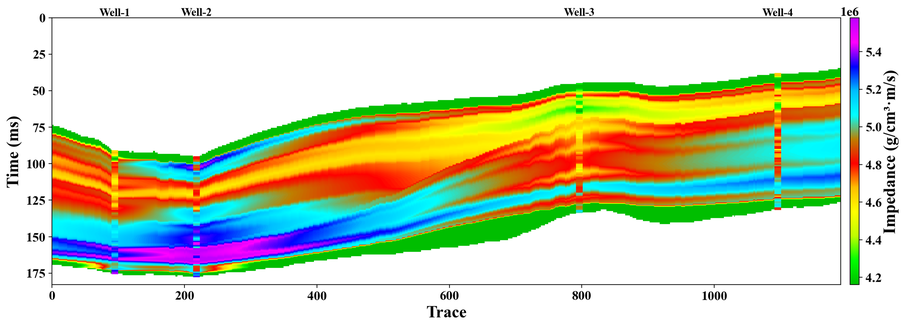}
        \caption{}
        \label{fig:c}
    \end{subfigure}

    \caption{\textbf{Field dataset 2 (F3 inter-well profile).} \textbf{(a)} Observed seismic profile. \textbf{(b)} Reference inversion result published with the dataset. \textbf{(c)} Associated low-frequency background impedance.}
    \label{fig:field_2}

    \vspace{-1em}
\end{figure}

\textit{Field dataset 2.}
To assess the robustness and transferability of the GeoVolDiff-generated pre-training data under a larger synthetic-to-field distribution gap, we experiment on an inter-well profile extracted from the F3 dataset (Fig.~\ref{fig:field_2}\subref{fig:a}). F3~\citep{dgb2009f3}, publicly released by dGB Earth Sciences, lies at the junction of the Step Graben and the Dutch Central Graben and exhibits complex structural and stratigraphic features that differ substantially from those in the synthetic training data. The proposed framework is compared with a reference inversion result published with the dataset (Fig.~\ref{fig:field_2}\subref{fig:b}) and its associated low-frequency background model (Fig.~\ref{fig:field_2}\subref{fig:c}).

\begin{figure}[h]
    \centering

    \begin{subfigure}[b]{1\linewidth}
        \centering
        \includegraphics[width=\linewidth]{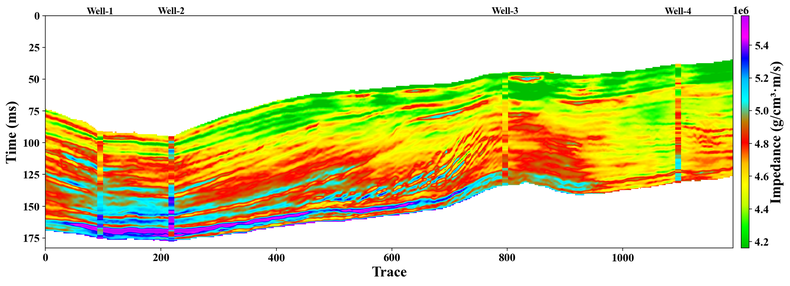}
        \caption{}
        \label{fig:a}
    \end{subfigure}

    \begin{subfigure}[b]{1\linewidth}
        \centering
        \includegraphics[width=\linewidth]{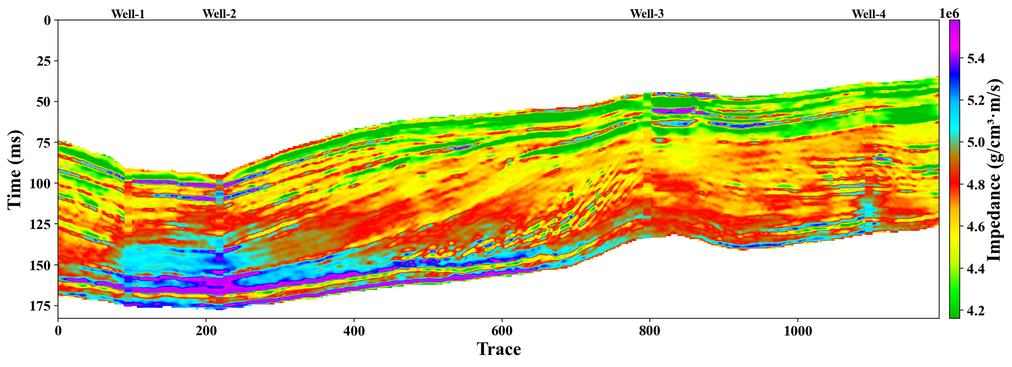}
        \caption{}
        \label{fig:b}
    \end{subfigure}

    \caption{\textbf{Inversion on the F3 profile with Ricker-wavelet pre-training.} \textbf{(a)} Pre-trained network without fine-tuning. \textbf{(b)} After fine-tuning with Well-1 to Well-4.}
    \label{fig:res_4}
    \vspace{-1em}
\end{figure}

Fig.~\ref{fig:res_4} reports the inversion results before and after fine-tuning. For a fair comparison with the reference, four well logs (Well-1 to Well-4, black) are used during fine-tuning. Even before fine-tuning, the pre-trained network produces a profile with strong lateral continuity; after fine-tuning, it yields physically reasonable impedance values without any low-frequency background model, indicating effective field-domain adaptation driven solely by sparse well-log supervision.

\begin{figure}[h] 
    \centering

    \begin{subfigure}[b]{1\linewidth}
        \centering
        \includegraphics[width=\linewidth]{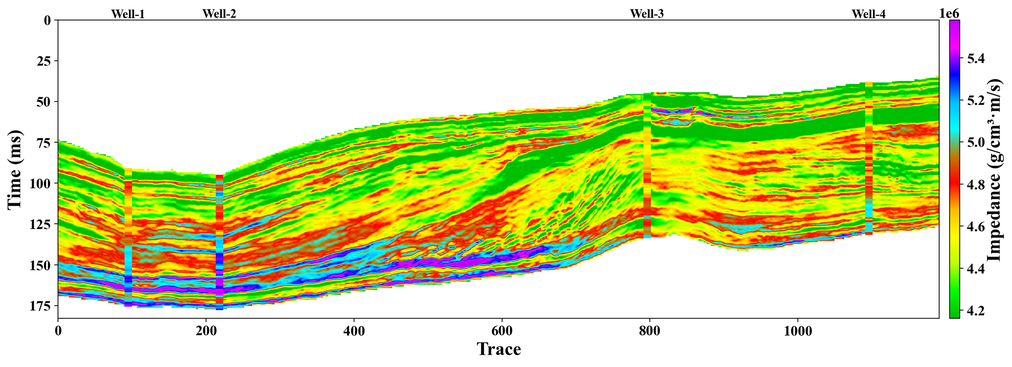}
        \caption{}
        \label{fig:a}
    \end{subfigure}

    \begin{subfigure}[b]{1\linewidth}
        \centering
        \includegraphics[width=\linewidth]{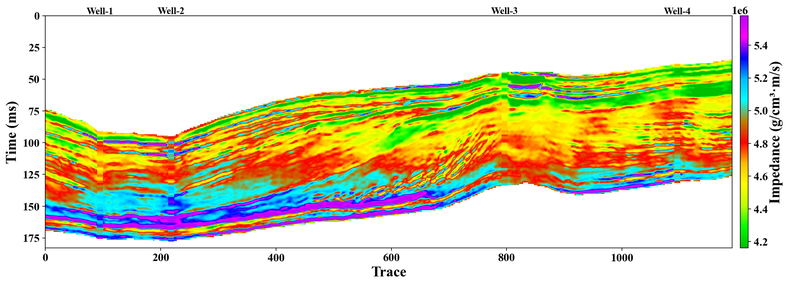}
        \caption{}
        \label{fig:b}
 
    \end{subfigure}

    \begin{subfigure}[b]{1\linewidth}
        \centering
        \includegraphics[width=\linewidth]{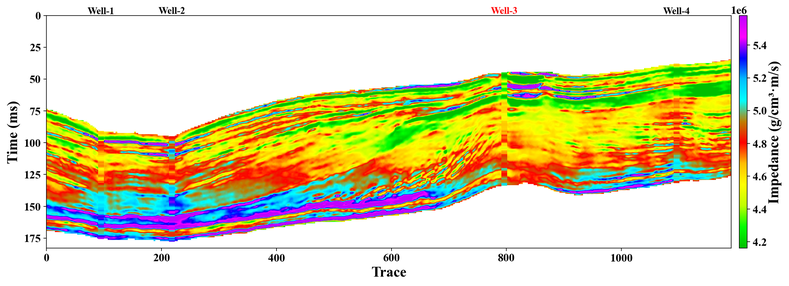}
        \caption{}
        \label{fig:c}
    \end{subfigure}

    \caption{\textbf{Inversion on the F3 profile with wavelet-adapted pre-training.} \textbf{(a)} Wavelet-adapted pre-trained network without fine-tuning. \textbf{(b)} After fine-tuning with Well-1 to Well-4. \textbf{(c)} Blind-well test with Well-3 (red) held out.}
    \label{fig:res_5}
    \vspace{-1em} 
\end{figure}

The field seismic wavelet differs substantially from the Ricker wavelet used at pre-training in phase, side-lobe structure, and dominant frequency. To mitigate this mismatch, the pre-training seismic data are re-synthesized by convolving the original impedance volumes with the field-estimated wavelet, and the network is then fine-tuned on this wavelet-adapted dataset. As shown in Fig.~\ref{fig:res_5}\subref{fig:a} and \subref{fig:b}, incorporating the field wavelet yields richer structural detail, and the fine-tuned result more closely matches the reference inversion in both resolution and impedance trend. A blind-well test is then carried out by holding out Well-3 (red) and fine-tuning on the remaining wells. As shown in Fig.~\ref{fig:res_5}\subref{fig:c}, the predicted impedance at the blind-well location agrees well with the withheld well-log measurements, further corroborating the generalisation of the pretrain–finetune framework.

Taken together, the results on both field datasets indicate that GeoVolDiff-generated data have sufficient geological plausibility and distributional coverage to serve as effective pre-training resources for downstream geophysical tasks, offering a viable route to alleviating the chronic scarcity of labelled training data in practical inversion workflows.

\section{Conclusion \& Discussion}
We have presented GeoVolDiff, a 3D geological volume generation framework that combines two complementary stages: a physics-based forward-simulation pipeline that constructs the initial training corpus, and a Latent Diffusion Model that supports both unconditional generation and fault-conditioned structural synthesis. Relative to forward simulation alone, the diffusion model enables more flexible and diverse data generation and offers a practical way to mitigate the chronic shortage of labelled training data in geophysical deep learning. We further validate the generated data as pre-training resources for seismic impedance inversion, providing empirical evidence of their geological plausibility and practical utility.

A broader observation from these results is that, under limited-sample conditions, access to sufficient and diverse training data may yield more direct performance gains than incremental refinements to network architectures or loss design. GeoVolDiff exploits a natural capability of diffusion models—data synthesis—to expand the geological training corpus. Unlike forward simulation, which requires considerable domain expertise and careful parameter selection, the trained diffusion model produces diverse, structurally coherent samples with minimal additional configuration. The generated 3D volumes exhibit strong lateral continuity and structural diversity (Sec.~3.2), and the downstream inversion experiments achieve competitive performance using only the pre-training corpus, without specialised architectures or elaborate loss formulations. This further supports the view that the synthesised samples carry geologically meaningful, representationally rich structure.

The current framework nonetheless has several limitations. Owing to GPU-memory constraints, the diffusion model is trained at $128^3$ voxels, below the $256^3$ native resolution of the forward-simulation pipeline—a constraint that does not arise in the 2D setting. Despite this gap, the present results are sufficient to establish the methodological feasibility of the framework. Looking ahead, GeoVolDiff can be deployed as a data-augmentation module under resource-constrained conditions and, as conditioning information becomes richer, can plausibly be extended into a full geological modelling methodology. We further emphasise that no real-field information—neither well logs nor seismic data—is used during diffusion-model training; field data enter only at the lightweight fine-tuning stage. This design encourages the pre-trained network to acquire generic structural representations and seismic-to-impedance mappings from large-scale synthetic data, which are then adapted to the target domain through sparse well-log supervision.

\begin{figure}[h]
    \centering
    \includegraphics[width=1\linewidth]{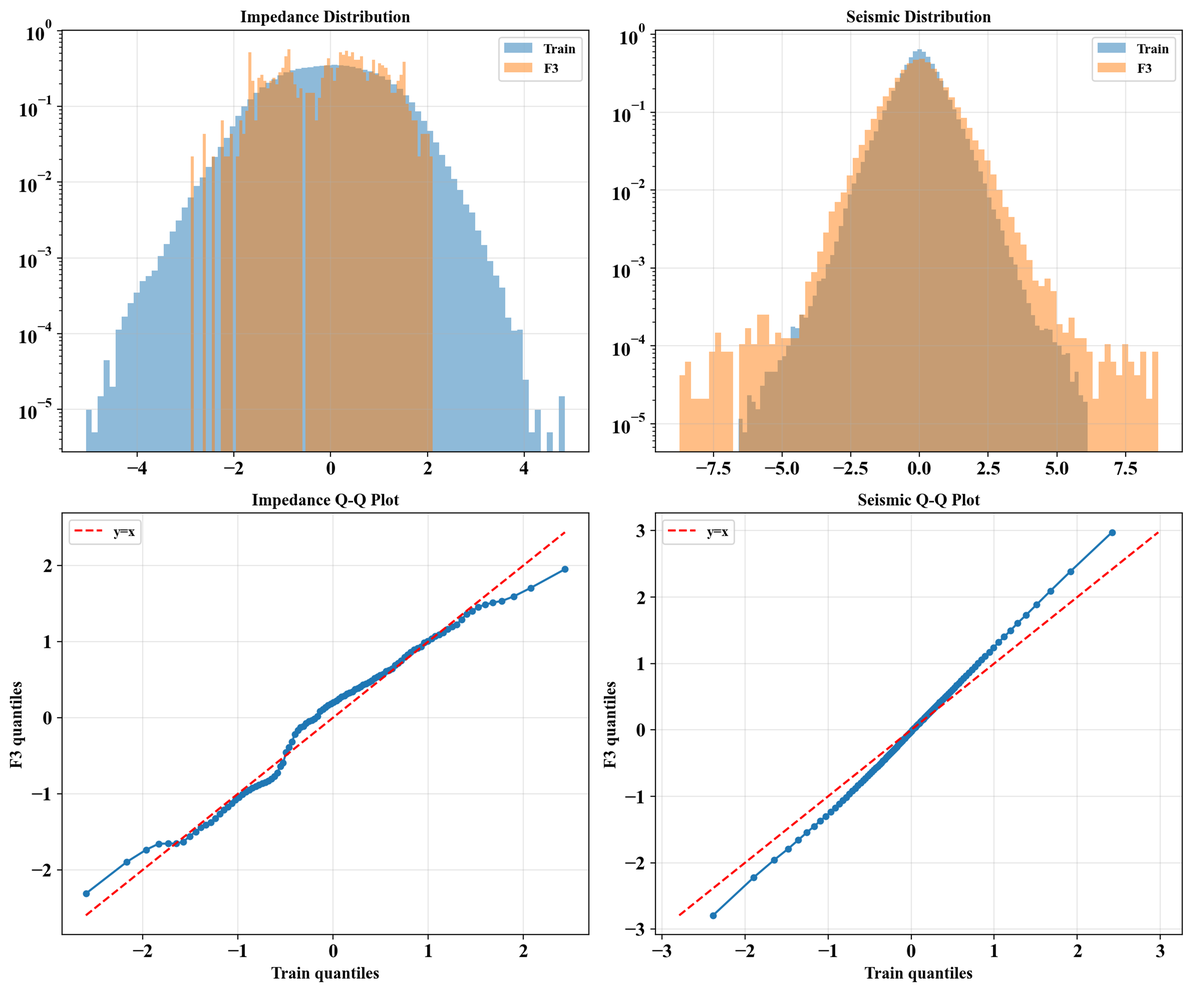}
    \caption{
        \textbf{Distributional comparison between pre-training data and F3 field observations.}
        Top row: histogram distributions of acoustic impedance 
        (left) and seismic amplitude (right) on a logarithmic 
        density scale. Bottom row: corresponding Q--Q plots.
    }
    \label{fig:qq}
    \vspace{-1em}
\end{figure}

To examine the distributional relationship between the pre-training data and the F3 field observations, we compare their well-log and seismic-amplitude distributions in Fig.~\ref{fig:qq}. The well-log distributions suggest that the pre-training data adequately cover the dynamic range of the available measurements. The seismic-amplitude distributions, however, indicate that the pre-training data only partially overlap with the real seismic distribution. We attribute this to two factors. \textit{First}, the subsurface impedance contrasts within the survey area span a broader range than the synthetic training set, producing a correspondingly wider amplitude range in the field-recorded seismic data. \textit{Second}, real seismic observations contain additional variability from complex wave-propagation effects, lateral geological heterogeneity, acquisition footprints, and processing artifacts, which the simplified convolutional forward modelling used for the synthetic data does not fully reproduce. One possible avenue to reduce the seismic-domain discrepancy is to replace the convolutional forward model with more realistic wave-equation-based seismic modelling. Then, at the geological level, a natural extension is to incorporate geological priors—such as low-frequency trends, interpreted structural frameworks, and regional geological constraints—as conditioning signals for the diffusion model. Such information can steer generation toward subsurface realizations more representative of a target geological setting, yielding structurally diverse yet geologically plausible models, potentially narrowing the gap between synthetic and real subsurface properties while enabling controllable geological model generation.

In summary, GeoVolDiff shows encouraging potential as a generative pipeline for 3D geological volumes. Non-trivial challenges remain on the path to real-world deployment—most notably training resolution, distributional coverage of seismic signatures, and richer structural conditioning—but the present results establish a scalable foundation for synthesis-driven geophysical deep learning and suggest that high-quality training data for downstream geophysical tasks can, to a substantial degree, be obtained from synthesised sources.

\newpage
\bibliography{reference}
\bibliographystyle{plainnat}
\newpage

\end{document}